\documentstyle[preprint,aps]{revtex}

\begin{document}
\draft
\preprint{UM-ChE-97/006}
\title{ Precise determination of the bond percolation thresholds and
finite-size scaling corrections for the s.c., f.c.c., and
b.c.c.\ lattices}
\author{Christian D. Lorenz and Robert M. Ziff}
\address{Department of Chemical Engineering,
University of Michigan, Ann Arbor, MI 48109-2136}

\date{\today}
\maketitle
\begin{abstract}

Extensive Monte-Carlo simulations were 
performed to study bond percolation on the simple cubic (s.c.),
face-centered cubic (f.c.c.), and body-centered cubic (b.c.c.) lattices,
using an epidemic kind of approach.  These simulations 
provide very precise values of the critical thresholds for each of
the lattices: 
$p_c($s.c.$) = 0.248\,812\,6 \pm 0.000\,000\,5$,
$p_c($f.c.c.$) = 0.120\,163\,5
\pm 0.000\,001\,0$, and
$p_c($b.c.c.$) = 0.180\,287\,5 \pm 0.000\,001\,0$.
For $p$ close to $p_c$, the results follow the expected
finite-size and scaling
behavior, with values for the Fisher exponent
$\tau$ $(2.189 \pm 0.002)$,
the finite-size correction exponent $\Omega$
$(0.64 \pm 0.02)$, and the scaling function
exponent $\sigma$ $(0.445 \pm 0.01)$
confirmed to be universal. 

\end{abstract}
\pacs{PACS numbers(s): 64.60Ak, 05.70.Jk}

\narrowtext

\section{Introduction}

Percolation theory is used to describe a variety
of natural physical processes, 
which have been discussed in detail by Stauffer and Aharony
\cite{staufferaharony}
 and Sahimi 
\cite{sahimi}.  In two-dimensional percolation,
either exact values or precise
estimates  are known for the critical thresholds
and other related coefficients
and  exponents \cite{sykesessam,ziffsuding,ziff92,nienhuis}.  

However, three-dimensional lattices are relevant
for most natural processes.  
The most common of these are the simple cubic (s.c.),
the face-centered cubic 
(f.c.c.), and the body-centered cubic (b.c.c.) lattices.  
The percolation thresholds  
for these lattices are not known exactly,
and the estimates that have been 
determined for the latter two lattices are much
less precise than the values that 
have been found for typical two-dimensional systems. 

An exception is the case of the s.c.\ lattice, where fairly precise
values have been determined.  A number of years ago,
Ziff and Stell carried out a study of three-dimensional
percolation for the site and bond
percolation on the s.c.\ lattice
\cite{ziffstell}. The values 
$p_c = 0.248\,812 \pm 0.000\,002$ and
$p_c = 0.311\,605 \pm 0.000\,010$ were found
for bond and site 
percolation, respectively.
The behavior for $p$ away from $p_c$ was also studied,
and it was 
found that the critical exponents can be described by the consistent set      
$\tau = 116/53 \approx 2.1887$, $\sigma = 24/53 \approx 0.4528$,
and $\nu = 7/8$,
with errors $\pm 0.001$, $\pm 0.001$, and 
$\pm 0.008$, respectively.  However, this work, which used a growth 
method essentially equivalent to the epidemic approach,
remained unpublished as an internal report only \cite{ziffstell}.  

In a more recent work, Grassberger found $p_c = 0.248\,814 \pm
0.000\,003$ and 
$p_c = 0.311\,604 \pm 0.000\,006$ for bond and site percolation
(respectively) on
the s.c lattice \cite{grassberger},
using an epidemic analysis \cite{grassdelatorre}.
The precise agreement between these two independent works provides 
stong evidence that the procedures,
random-number generators, error analyses,
etc., implemented by both of these groups of authors are correct.
Previous to these two works, $p_c$ was known only to about four significant
figures, as discussed below.
It is important to know $p_c$ very precisely so that other
critical properties can be studied without bias.  

For bond percolation on the f.c.c.\ and b.c.c.\ lattices,
the most accurate
values  appear to be $p_c = 0.1200 \pm 0.0002$ 
and $p_c = 0.1802 \pm 0.0002$, respectively,
recently found by van der Marck \cite{vandermarck} using the 
average crossing probability method \cite{staufferaharony}.
Additional literature values 
are  listed below.  These thresholds are not precise enough for
a study we have been carrying out on the universal
excess cluster numbers in 3-D systems
\cite{lorenzziff}, and as a consequence we have conducted the
present work.

Here, we use a growth or epidemic analysis to obtain new,
high-precision values for the percolation thresholds of 
the s.c., f.c.c., and b.c.c.\ lattices.
We reaffirm that three-dimensional
percolation follows the hypothesized 
finite-size and scaling behavior, and estimate the exponents and 
coefficients that enter into the finite-size and scaling functions.

In the following sections, we discuss the simulation
that we used to grow the 
percolation clusters and obtain our data.
Then, we present and briefly discuss the results that we 
obtained from our simulations.

\section{Simulation Method}

We used a Monte-Carlo simulation of bond percolation on each of the 
three-dimensional lattices.  
This simulation employed the so-called Leath growth 
algorithm \cite{leath} to generate individual percolation clusters.  
The cluster was
started  at a seeded site that was centrally located on the lattice. 
From this site, a 
cluster was grown to neighboring sites by 
occupying the connecting bonds 
with a certain probability $p$ or leaving them 
unoccupied with a probability 
$(1-p)$.  The unit vectors that were used to 
locate the neighboring sites for each 
lattice are summarized in Table I. \  
Each of these clusters were allowed to grow 
until they reached an upper cut-off, 
at which point any cluster that was still 
growing was halted.  This cut-off was 
$2^{21}$ (2,097,152) wetted sites for the s.c.\ 
lattice and $2^{20}$ (1,048,576) wetted 
sites for the f.c.c.\ and b.c.c.\ lattices.  
The size of the cluster was characterized 
by the number of wetted sites rather
than the  occupied bonds of the clusters, 
since the former figures directly in
the  growth algorithm procedure.

In order to grow such large clusters, 
our simulation used the
data-blocking scheme \cite{ZCS}, 
in which large lattice is divided
into smaller blocks and memory is assigned to a block only
after the cluster has grown into it.
 In this case, the lattice, which has dimensions of 
$2048 \times 2048 \times 2048$, was 
divided into $64^3$ blocks of dimensions 
of size $32 \times 32 \times 32$.
Bit mapping was also used to reduce the
memory load  of the large lattices.  The upper six 
bits of each coordinate were used to tell 
where in the memory that block is 
mapped.  The lower five bits of the 
$y$- and $z$-coordinates tell the memory 
which word and the lower five bits of the 
$x$-coordinate tell the memory which 
bit of that word is used to store the site.  
With such a large lattice and the cut-offs
we  used, none of the clusters saw the lattice boundary, 
so there were no boundary 
effects.

The simulation counted the number of clusters 
that closed in a range of $(2^n, 
2^{n+1} - 1)$ sites for $n = 0, 1, \ldots $, 
and recorded this number in the $n$th 
bin.  If a cluster was still growing 
when it reached the upper cut-off,  
it was counted in the  last (20th or 21st) bin.  
The simulation also kept track of the 
average number of occupied bonds and 
the average number of unoccupied 
bonds for clusters in each bin range, 
for reasons discussed below.

Random numbers were generated by the four-tap shift-register rule
$x_n = x_{n-471} \oplus x_{n-1586} \oplus x_{n-6988} \oplus x_{n-9689}$
where $\oplus$ is the exclusive-or operation, which we have used
(with apparent success) in numerous 
previous studies (e.g., \cite{ziffsuding,ziff92}).

\section{Results}
\subsection{Fisher exponent $\tau$}

Using the data obtained from the simulation,
the number of clusters grown to 
a size greater than or equal to size $s$ can be deduced.   
The probability of growing clusters with
the number of wetted sites $\ge s$,
$P(s,p)$, when operating at the critical 
threshold ($p = p_c$) and neglecting finite-size effects, 
is predicted to follow 
\cite{staufferaharony}:
\begin{equation}
P(s,p_c) \sim A s^{2-\tau}
\label{eq1}
\end{equation}
In Fig.~1, data from our simulation plotted on a log-log plot
 shows good agreement with
this equation.   However, for small clusters, 
there is a clear nonlinear region, which
represents the  finite-size effect.  
For large $s$, the curves also become nonlinear, 
resulting from growing the clusters at values of $p$ away from $p_c$.
Values of $p$ that are greater than $p_c$
 produce curves that increase for large $s$, and values 
of $p$ below $p_c$  cause curves to decrease. 

Such behavior is similar to that seen in an epidemic
analysis of a transition of interacting particle
systems, and indeed, the cluster growth procedure
is in fact essentially such a 
process.  The only difference is that, in the usual epidemic
analysis \cite{grassdelatorre}, one plots the growth probability
as a function of the time (generation or chemical distance number
for percolation), while here we plot that probability as a function
of the size.  But since the size scales with time, the two 
approaches are equivalent.

The slope of the linear portion of the curves 
shown in Fig.~1 is equal to ($2-\tau$),
where $\tau$ is the Fisher exponent
\cite{fisher}.  However, due to the nonlinear 
portions of the  curves and the uncertainty
of which $p$ is precisely at the critical
threshold,  it is difficult to accurately 
deduce $\tau$ from these plots.
To make the behavior more pronounced, we plot
difference between the data and a straight line of slope $2-\tau$
in Fig.~2, for the f.c.c.\ lattice and different values of $p$.
Here, the correct value of
$\tau$ produces a horizontal central portion of the curve.
Applying this analysis for all
three lattices yields $\tau  = 2.189 \ \pm 0.002$.
Fig.~3 shows the effect of using slightly larger 
or smaller values of $\tau$ for the 
f.c.c.\ lattice.

\subsection{Finite-size corrections.}

In order to account for the small clusters, 
the finite-size correction must be 
added to (\ref{eq1}).  Then $P$ is
predicted to follow
\begin{equation}
P(s,p_c) \sim A s^{2-\tau}(A + Bs^{-\Omega} + \ldots)
\label{eq2}
\end{equation}
where $\Omega$ is the first correction-to-scaling exponent \cite{hoshen}.  
Like (\ref{eq1}), (\ref{eq2})  is only valid at the critical threshold.  
At the correct value of $\Omega$,  (\ref{eq2})  predicts there will be
a linear relationship between $s^{\tau-2}P(s, p_c)$ and $s^{-\Omega}$.
Fig.~4  shows plots between these two quantites
for the s.c., f.c.c., and b.c.c.\ lattices, respectively.  
The values of $\Omega$ which produced the best linear 
fit are summarized in Table II. 
The slope of the curves gives the value for the 
coefficient $B$ in (\ref{eq2})  
and the intercept gives $A$ in (\ref{eq1})  and (\ref{eq2}).  
The values we found are summarized in Table III.   
Note that the last data point is shown on each of 
these plots, but they have been left out of linear 
fits because they represent clusters 
of only two sites, too small to be describe by just
the first term of the finite-size  corrections series.

\subsection{ Percolation thresholds.}
In Fig.~1, we show the effect that values of
$p$ away from $p_c$ have on the data
for  large clusters.  In order to account for
the behavior when $p \ne p_c$, a
scaling  function needs to be included in $P$.
The behavior is then represented by
\begin{equation}
P(s,p_c) \sim A s^{2-\tau}f((p-p_c)s^\sigma)
\label{eq3}
\end{equation}
which is valid rigorously in the scaling limit as
$s \to \infty$, $p \to p_c$, such that
 $(p - p_c)s^{\sigma}$ = constant \cite{staufferaharony}.
Because $p$ is close
to $p_c$, we can expand $f(x)$ in a Taylor series:
\begin{equation}
f((p-p_c)s^{\sigma})\sim 1 + C(p-p_c)s^\sigma + \ldots .
\label{eq4}
\end{equation}
Eqs.~(\ref{eq3}) and  (\ref{eq4}) describe how $s^{\tau-2}P(s, p)$
deviates from a constant value 
for large $s$ when $p$ is close to $p_c$.
Figs.~5(a), 5(b), and 5(c) show the plots of $s^{\tau-2}P(s, p) $
vs.\ $s^\sigma$ for the s.c., f.c.c.,
and b.c.c.\  lattices, respectively. For
these plots, we used  the value of
$\sigma = 0.453$ from \cite{ziffstell}, and
the linearity of the plots confirms that it is a  good value.
These plots show
a steep decline, which is the finite-size effect, 
followed by a mostly linear
region as predicted by (\ref{eq4}). 
The linear portions of  the curve become
more nearly horizontal as the chosen value
of $p$ gets closer  to $p_c$.  The
value for the critical threshold can
be deduced by interpolating from  these
plots.  For the three lattices considered here, we thus find:
\begin{mathletters}
\label{generalpclabel}
\begin{eqnarray}
		&&p_c(\hbox{s.c.}) = 0.248\,812\,6 \pm 0.000\,000\,5  \\
		&&p_c(\hbox{f.c.c.}) = 0.120\,163 5 \pm 0.000\,001\,0   \\
		&&p_c(\hbox{b.c.c.}) = 0.180\,287\,5 \pm 0.000\,001\,0
\end{eqnarray}
\end{mathletters}
A more formal method of carrying out this
interpolation can be obtained by 
plotting the slopes of the curves,
shown in Figs.~5(a), 5(b), and 5(c), 
vs.\ $p$. From (\ref{eq3})  and (\ref{eq4}),
it follows that:
\begin{equation}
{\partial(s^{\tau-2}P(s,p)) \over \partial (s^\sigma)}
= C(p-p_c) + \ldots
\end{equation}\label{eq5}
which implies that the value of the critical
threshold can be calculated from the
$x$-intercept  of this plot, as shown in Fig.~6.
Figs.\ 5(a) and 6 yield consistent values of 
the critical threshold for the s.c.\ lattice.     

\subsection{Scaling function.}
The linearized scaling function, as shown in
(\ref{eq3})  and (\ref{eq4}), can be 
studied efficiently by the following procedure.
 From (\ref{eq3})  and
(\ref{eq4}), it follows  that:
	\begin{equation}
{\partial P(s,p) \over \partial p} = AC s^{2-\tau + \sigma} + \ldots
\label{eq6}
\end{equation}
Eq.~(\ref{eq6})  shows that we could determine $2-\tau + \sigma$ and
$AC$ directly from a measurement of the derivative
$\partial P / \partial p$ as a
function of $s$.

To develop a formula for this derivative,
we consider the formal expression for 
the probability of growing a cluster containing
$n$ occupied bonds and $t$ vacant
(perimeter) bonds:
\begin{equation}
P_{n,t} = g_{n,t} p^n (1-p)^t
\label{eq7}
\end{equation}
where $g_{n,t} $ is the number of distinct clusters configurations
with $n$ occupied and $t$ vacant bonds.  From (\ref{eq7}),
we can write the probability of growing a cluster
of size greater than or equal
to
$s$ as  
	\begin{equation}
P(s,p) = \sum_n \sum_t g_{n,t} p^n (1-p)^t
\label{eq8}
\end{equation}
where the sum is over all $n$ and $t$ such that
the number of wetted sites is 
$\ge s$.

Differentiating (\ref{eq8}) with respect to $p$ gives: 
\begin{equation}
{\partial P(s,p) \over \partial p} = \sum_n \sum_t g_{n,t} p^n (1-p)^t
\left( {n\over p} - {t\over 1-p} \right)
\end{equation}\label{eq9}
which can be simplified to the final form:
\begin{equation}
{\partial P(s,p) \over \partial p} = 
{\langle n \rangle \over p} - {\langle t \rangle \over 1-p} ,
\label{eq10}
\end{equation}
where $\langle n \rangle$ and  $\langle t \rangle$
are the average number of
occupied and unoccupied   bonds, respectively,
over all clusters of size  $\ge
s$.   For a Monte-Carlo estimate of this derivative,
we simply use
$\langle n \rangle$ and  $\langle t \rangle$ averaged
over the sample of clusters.
The values for these averages were 
recorded by our simulations, and were then used with (\ref{eq10}) to 
calculate the derivative.

Fig.\ 7 shows a plot of the measured derivative (\ref{eq10})
vs.\ $s$ for the f.c.c.\ and b.c.c.\ 
lattices.  From the slopes and intercepts, and using the values
 of $A$ and $\tau$ determined above,
 we can deduce $\sigma$ and $C$.  For $\sigma$ we find 
$0.445 \pm 0.01$ for both cases, which is 
consistent with \cite{ziffstell}, and the
values of $C$ we find are given in Table III.

Another application of the derivative (\ref{eq10})
is that it can be used to
produce new  curves at values of $p$ near $p_c$ that
resemble the curves seen in
Figs.~5(a),  5(b), and 5(c), by virtue of a Taylor expansion,
\begin{equation}
P(s,p_2) = P(s,p_1) + (p_2-p_1) {\partial P(s,p) \over \partial p}
\Bigg|_{p=p_1}
\label{eq11}
\end{equation}
where the last derivative is determined using (\ref{eq10}).
This `shifting' of the data is useful for correcting results
taken close to, but not precisely at, $p_c$,
without carrying out a new set of time-consuming simulations.

\section{Discussion of Results}

\subsection{Fisher exponent $\tau$}

Our results show excellent agreement with
the expected relationship for the 
number of cluster of size greater than or equal to $s$,
which is shown in (\ref{eq1}).  The  only nonlinear portions
of the curve on a log-log plot occur when $s$ is
either small or large, which is where (\ref{eq1})
is not valid.  These nonlinearities are caused by
the finite-size effect (small
$s$) and the departure of $p$ from $p_c$ (large $s$).  

The value of $\tau$ was found by comparing curves
that were similar to those shown in Figs.\ 2 and 3,
finding the value that produced the most nearly
 horizontal curve.  As can be
seen in Figs.\ 2 and 3 for the f.c.c. lattice,
the reported  value of $\tau = 2.189$ \cite{ziffstell}
gives the best horizontal curve.  This value of $\tau$
also  provides the best horizontal curves for the other
two lattices, showing universality (with error bars of
$\pm 0.002$). This value of $\tau$ is also
consistent with the  values given by Stauffer and Aharony (2.18)
\cite{staufferaharony} and Nakanishi and Stanley
(2.19) \cite{nakanishi}.  

\subsection{Finite-size corrections.}

In Fig.\ 4, our data show excellent agreement
to the prediction described by 
(\ref{eq2}). \ It is expected that the 
exponent $\Omega$ is a universal quantity while
the coefficients $A$ and $B$ 
are different for the three different lattices.
These expectations were 
born out by our results showing the best fit of
data for all three lattices 
occurred at approximately the same value of $\Omega$ $(0.64 \pm 0.02)$,
as shown in Table II, but different values
of the coefficients as given in Table
III.  Our value  for $\Omega$
is significantly larger than the value reported by
Nakanishi and Stanley  ($\Omega = 0.40$) \cite{nakanishi}. 
We believe this discrepancy is due to their using less precise values of 
the critical exponents, which leads to more inherent error in $\Omega$,
and carrying out significantly 
fewer simulations, as were feasible at the time that that work was done.   

\subsection{Percolation thresholds.}

In Figs.\ 5(a), 5(b), and 5(c), our data show
excellent agreement with the 
predicted relationship as described by
(\ref{eq3})  and (\ref{eq4}).  The curves
show  the finite-size effects for small $s$ and then become linear.
The linear portion  of the curves become more nearly
horizontal as the value of
$p$ approaches $p_c$,  which is also predicted by the equations.

Our values for the critical thresholds
(\ref{generalpclabel}) are 
consistent with most previous works.
The critical threshold for the f.c.c.\ 
lattice has been reported as
$0.1200 \pm 0.0002$ \cite{vandermarck} and 0.119 
\cite{staufferaharony}.  The 
critical threshold for the b.c.c.\  lattice has been reported as
 $0.1802 \pm 0.0002$ \cite{vandermarck} and 
0.1803 \cite{staufferaharony}.
 The critical threshold for the s.c.\ lattice has been 
reported as $0.248\,812 \pm 0.000\,002$ \cite{ziffstell},
$0.248\,814 \pm 0.000\,003$ \cite{grassberger},
0.2488 \cite{staufferaharony},
$0.24875  \pm 0.00013$ \cite{grass86},
$0.2488 \pm 0.0002$ \cite{adler}, and 
$0.2487  \pm 0.0002$ \cite{vandermarck}.
A number of years ago the higher values
$0.2492 \pm 0.0002$ \cite{wilke}
and $0.2494 \pm 0.0001$ \cite{zabolitzky} were reported, but these have
since been recognized as probably erroneous due to
a flaw in programming \cite{SAA}.

\subsection{Scaling function.}

One expects that the value of the scaling function
exponent $\sigma$ should be a 
universal quantity while the coefficient $C$ should
be different for each of the 
lattices.   Values of these quantities were only
calculated for the f.c.c.\ and
b.c.c.\ lattices  because in the s.c.\ lattice
simulations we did not record the
values of $\langle n \rangle$ and 
$\langle t \rangle$.  The same value of
$\sigma$ ($0.445 \pm 0.01$) produced the best
fit for both lattices.   This
value is significantly smaller than the value
reported by Nakanishi and  Stanley
($0.504 \pm 0.030$) \cite{nakanishi},
but it is  consistent with the values
given by Ziff and Stell \cite{ziffstell} ($0.453
\pm 0.001$) and  Stauffer and Aharony \cite{staufferaharony} (0.45).
The values for the coefficient $C$ which
are summarized in Table III, were different for the two lattices.
Thus, these results  confirm the expectations.

\section{Conclusion}

Our work has produced the bond percolation
critical threshold values given
in (\ref{generalpclabel}). For the f.c.c.\ and
b.c.c.\ lattices these results are at least
two orders of magnitude more precise
than previous values, while for the s.c.\ lattice the result is four times
more precise.
We find critical exponents consistent with those given in
\cite{ziffstell}, and a more precise value
of the correction-to-scaling exponent $\Omega$ than previously known.
This increased
precision is the result of having conducted extensive simulations
with an inherently efficient procedure
(the epidemic method
\cite{grassdelatorre}) along with programming
techniques that allow one to
simulate very large lattices \cite{ZCS}. The results of this work 
allow other properties of three-dimensional percolation on these
lattices to be studied equally precisely. 

Because all of this work was performed
in a relatively short amount of time, 
we believe that the epidemic approach is a more efficient way to find the 
critical threshold than the conventional crossing-probability methods.
However, we did not make a direct time comparison of two  methods.
Note that the epidemic growth algorithm used here can also be used to
study systems of  any dimension, unlike hull methods
\cite{ziffsuding,rosso,ziffsapoval}, which are limited to two dimensions.

Our measurements of $P(s,p)$ for all $s$ (except the smallest values)
and for $p$ very close to $p_c$ can be summarized in a
single equation 
\begin{equation}
P(s,p) = s^{\tau - 2} (A + B s^{-\Omega} ) (1 + C (p-p_c) s^\sigma )
\label{eq12}
\end{equation}
with all constants and exponents determined by our simulations.
Although this equation mixes finite-size and bulk scaling forms,
it provides a very  accurate fit of the data in the regime we studied.

\acknowledgements{This material is based upon work supported by the US
National Science Foundation under Grant No.\thinspace DMR-9520700}

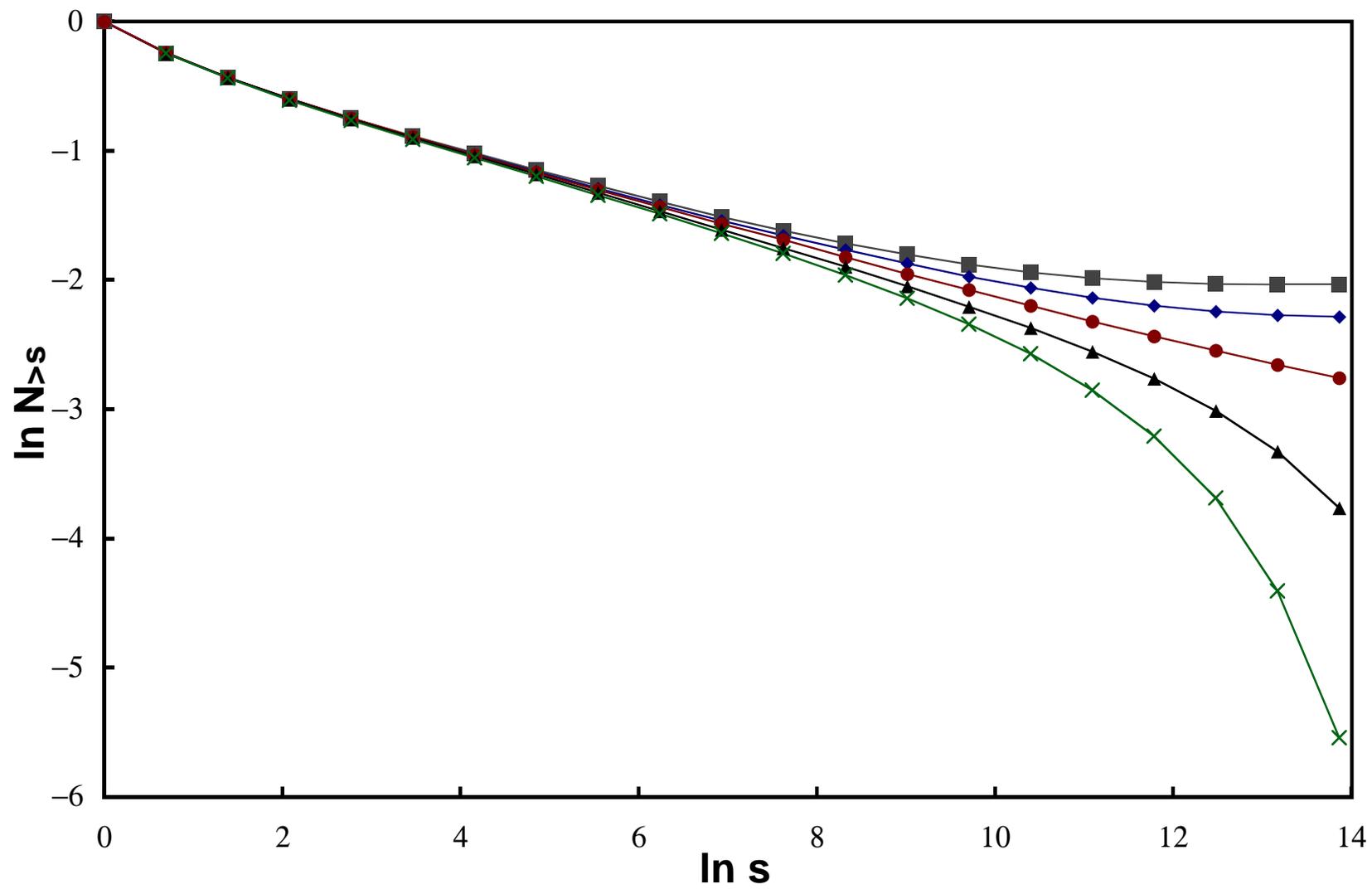
\begin{figure}
\caption{ Plot of the raw simulation results on a log-log plot.  This data is
from  the f.c.c.\ lattice and each of the five curves represent a different
value of $p$:  0.1206 (square), 0.1204 (diamond), 0.1202 (circle), 0.1200
(triangle), and 0.1198  (cross).  Nonlinearities are caused by finite-size
effects (small $s$) and $p$ not equalling $p_c$ (large $s$). }
\end{figure}

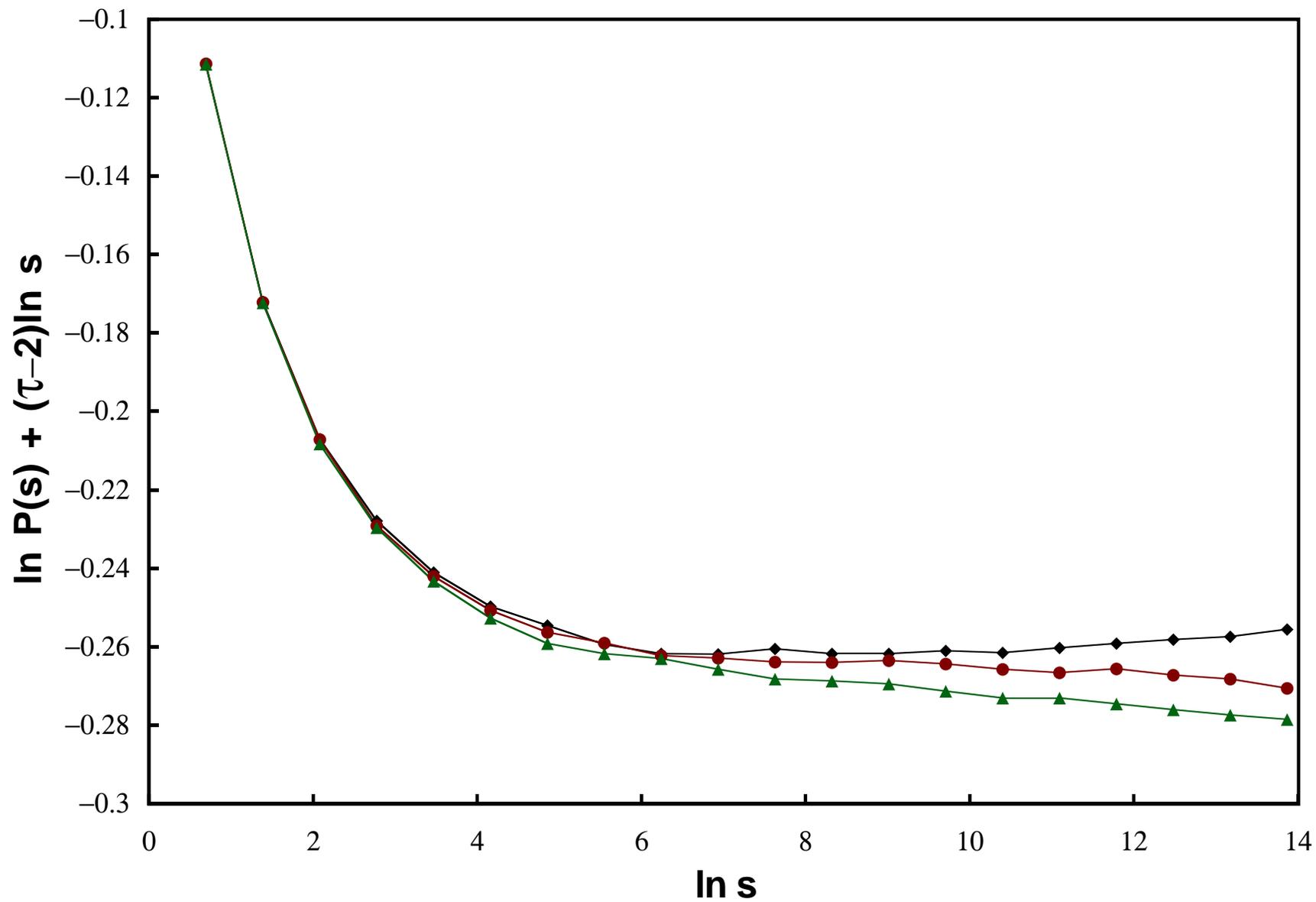
\begin{figure}
\caption{ Determination of the Fisher exponent, $\tau$,
by plotting ln $s^{\tau-2}P(s, p_c)$ vs.\ 
ln $s$.  These curves show data from the simulations of a
f.c.c.\ lattice and each curve represents a different value
of $p$: $0.120\,165$ (diamond), $0.120\,162$ (circle),  and $0.120\,160$
(triangle).  These curves were produced using a value of $\tau$ =  2.189,
which produced the best horizontal curve for all three lattices.}
\end{figure}

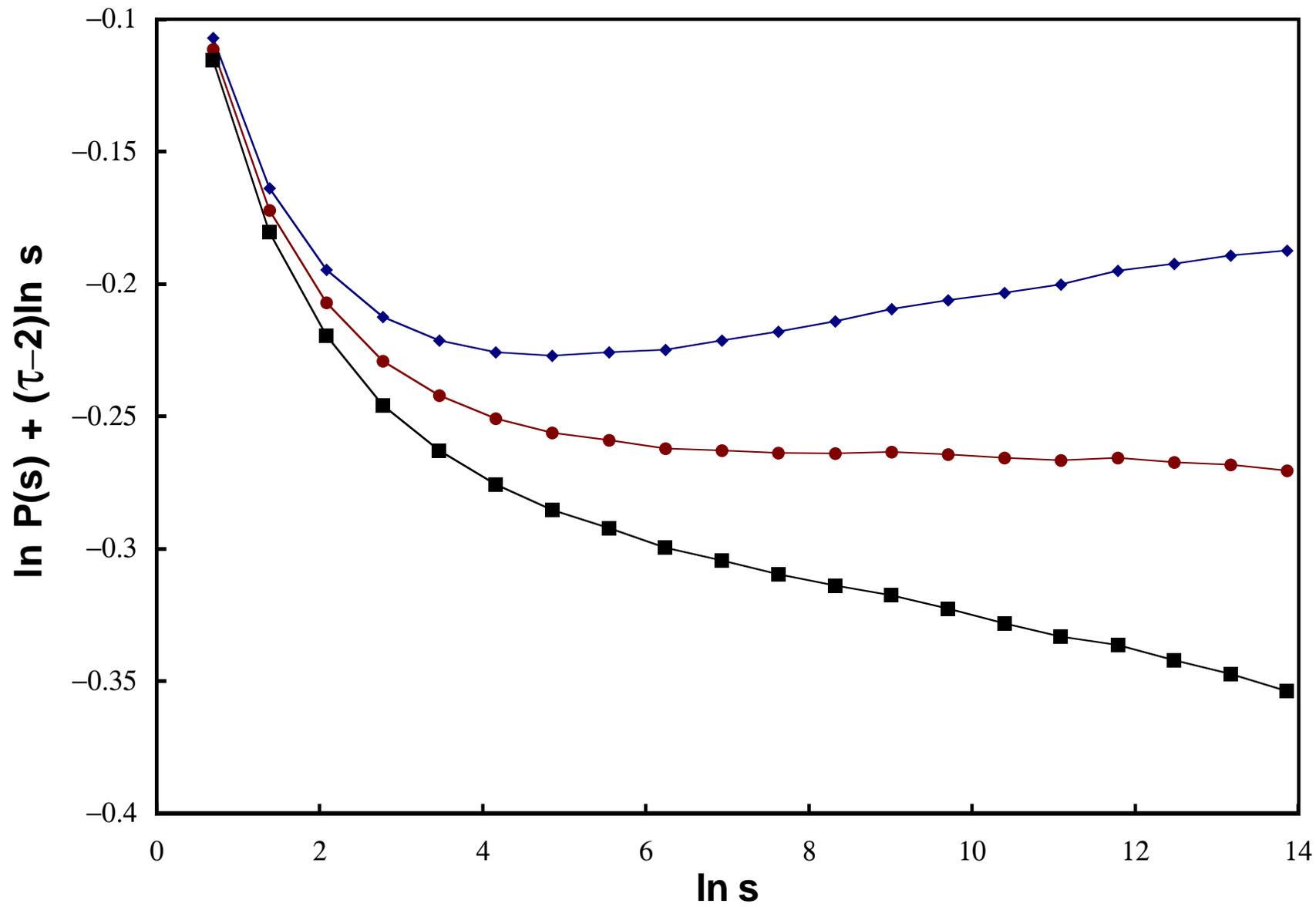
\begin{figure}
\caption{Effect of varying $\tau$ on the central curve of Figure 2
(f.c.c.\ lattice).   These curves (all at $p = 0.120\,162$, which is close to
$p_c$) compare three different  values of $\tau$: 2.192 (diamond), 2.189
(circle), and 2.186 (square). The best horizontal fit is clearly produced by
usingthe middle value.}
\end{figure}

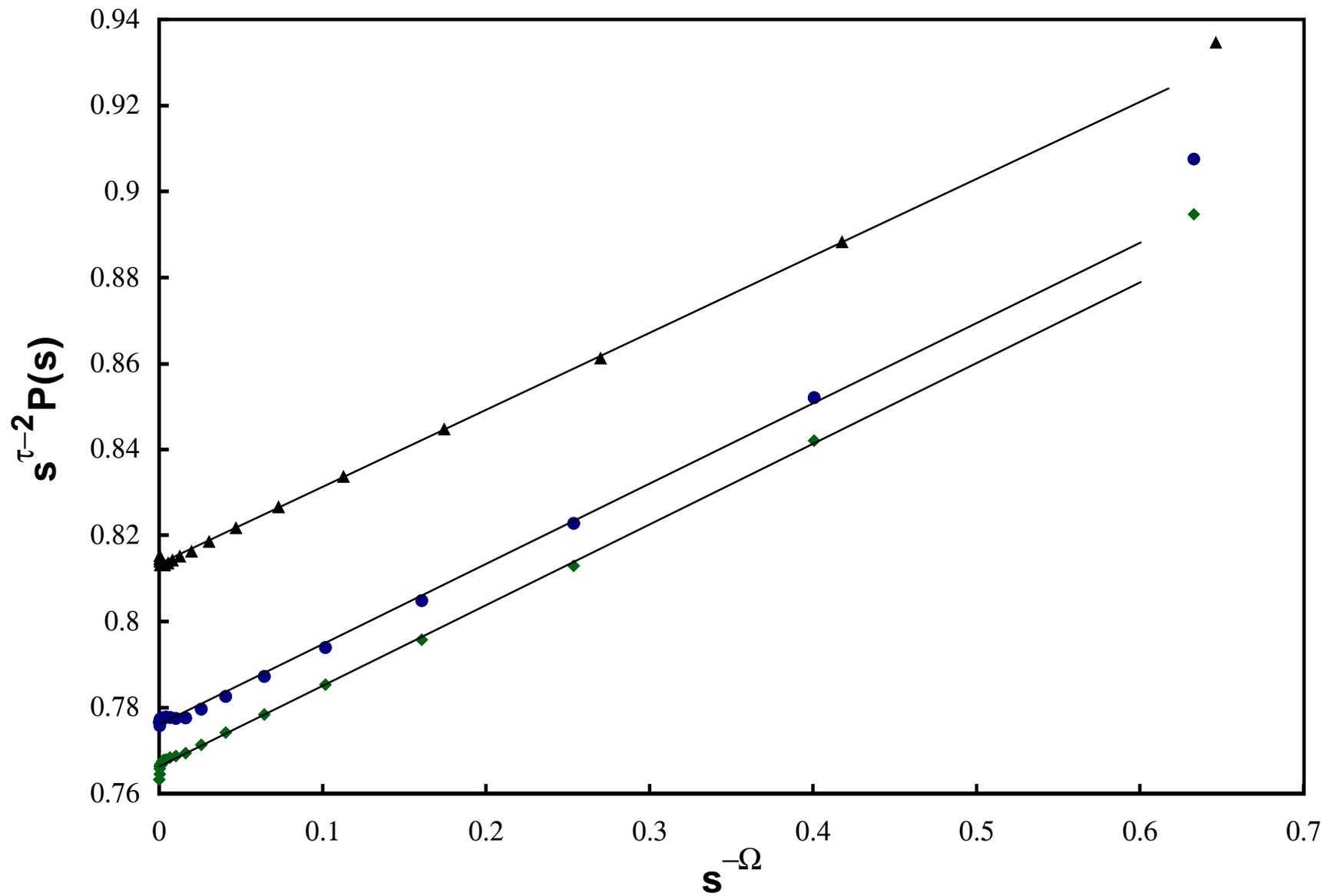
\begin{figure}
\caption{ Plot to determine the finite-size correction predicted by (2), 
for the s.c.\ (triangle), b.c.c.\ (circle), and f.c.c.\ (diamond) lattices.
The values of $\Omega$, $A$, and $B$ are summarized in Tables II and III.
The last points of each curve, which represented clusters of just
two wetted sites, were left out of the curve fit.  Deviations from linearity
for large $\tau$ are caused by $p$ being slightly different  from $p_c$.}
\end{figure}

\begin{figure}
\caption{Plot of the data vs.\ $s^\sigma$.
The linearity for large $\tau$ demonstrates the 
validity of (3)  and (4). The value of $p$ that produces the best 
horizontal fit yields the critical threshold.
Figure 5(a) shows the data from the 
s.c.\ lattice, and each curve shows a different
value of $p$: $0.248\,820$ (diamond), 
$0.248\,814$ (square), $0.248\,810$ (circle),
and $0.248\,800$ (triangle).  Figure 5(b) shows 
the data from the f.c.c.\ lattice,
and the values of $p$ shown are: $0.120\,165$
(diamond), $0.120\,162$ (circle), and $0.120\,160$ (triangle).
Figure 5(c) shows the data from the b.c.c.\  lattices,
and the values of $p$ are: $0.180\,300$ (diamond), $0.180\,
287\,5$ (circle), and $0.180\,275\,0$ (triangle).}
\end{figure}

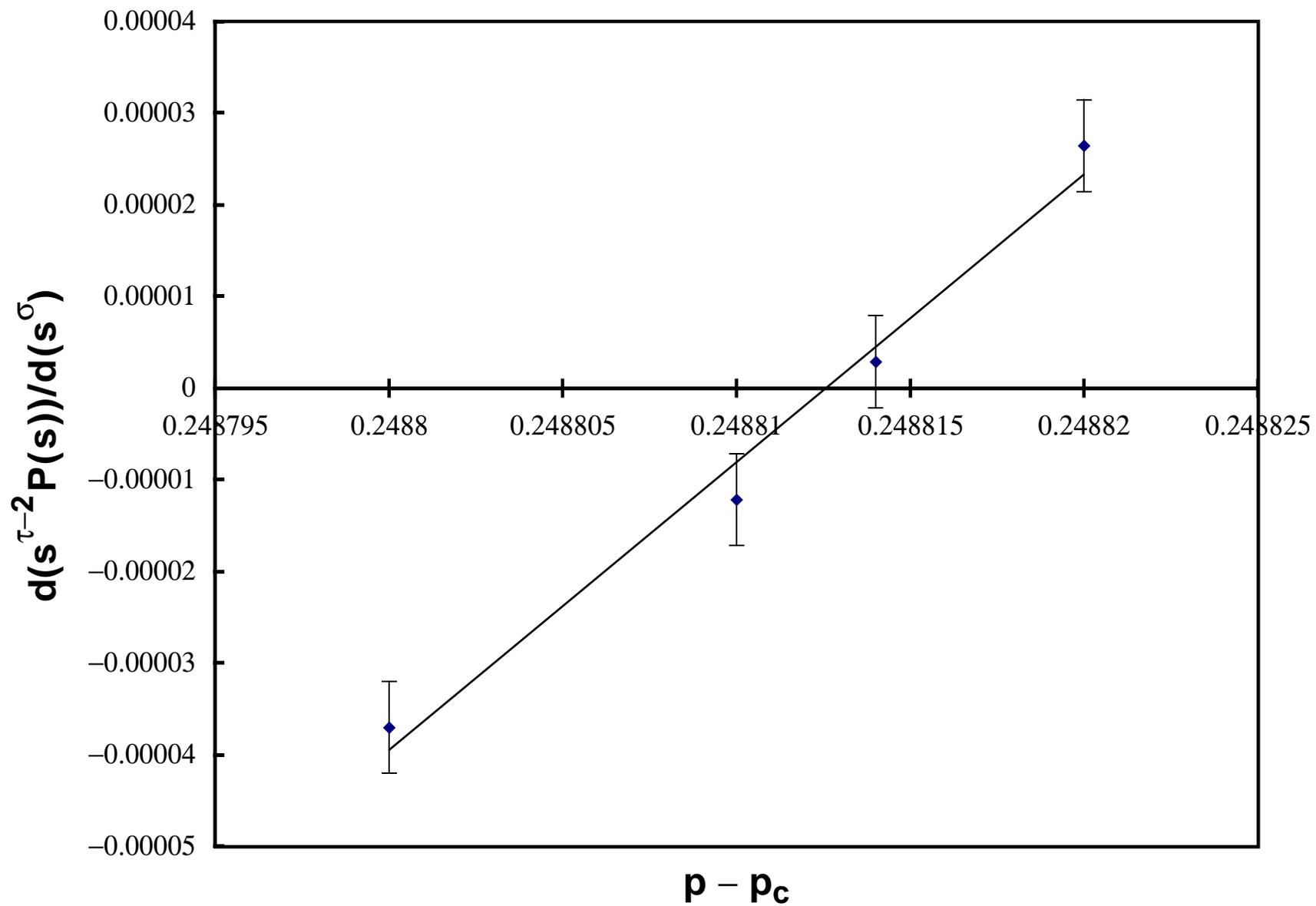
\begin{figure}
\caption{Plot of the slope of the linear portion of the curves shown in Figure 
5(a) as a function of $p$.  The $x$-intercept gives the value of the critical 
threshold for the s.c.\ lattice.}
\end{figure}

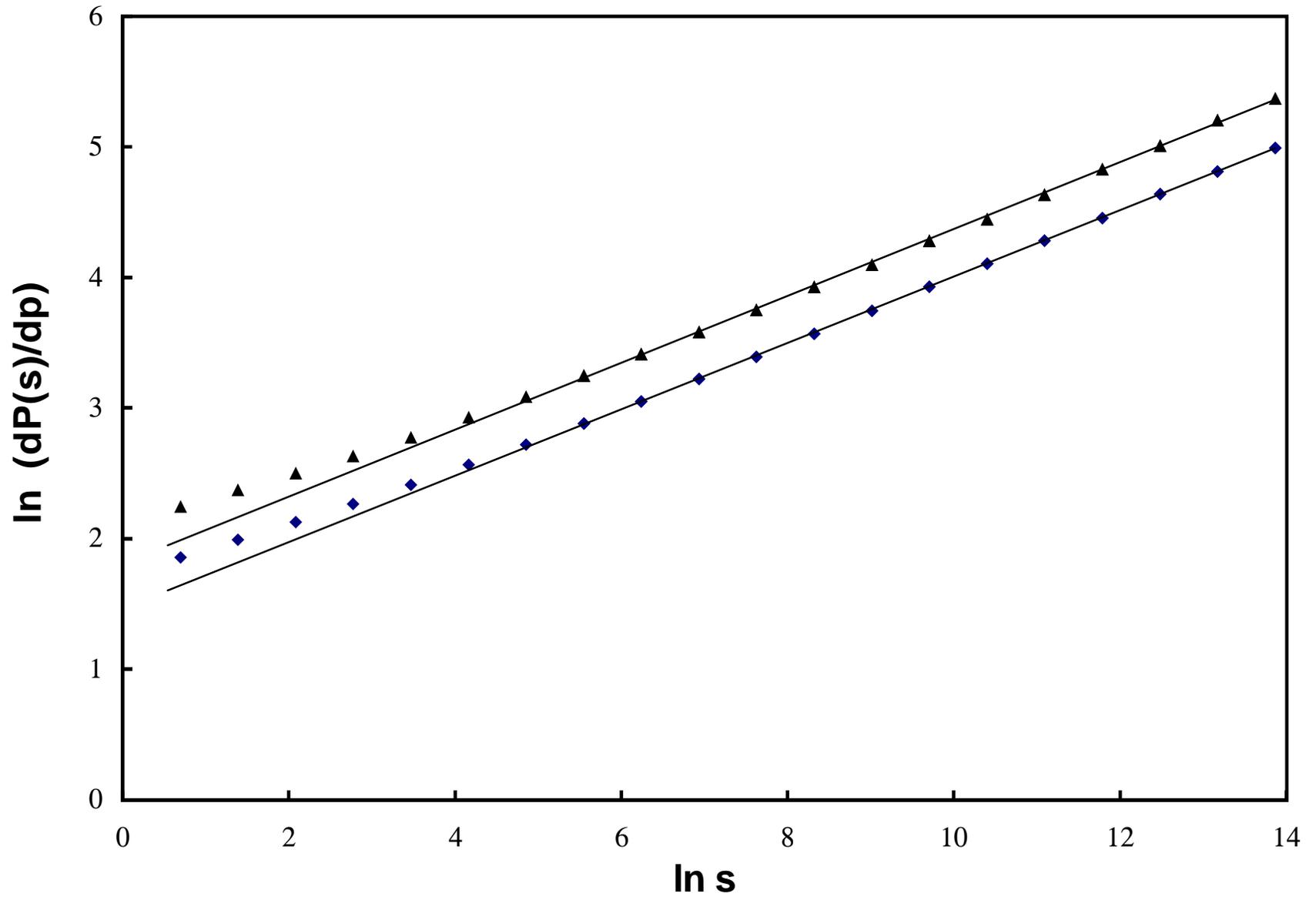
\begin{figure}
\caption{ Determination of $\sigma$ and $C$ using $\partial P/\partial
p$ calculated from (11). By fitting the last fifteen data  points to (7),
$2-\tau+\sigma$ can be found from the slope of these curves  and $AC$ can be
found from the $y$-intercept. The resulting values for $\sigma$ and 
$C$ (using $\tau$ and $A$ determined above) for the f.c.c.
(triangle) and b.c.c.\ (diamond) lattices are summarized in Tables II and III,
respectively.}
\end{figure}

\begin{table}
\caption{Unit vectors used to describe the neighbors in the s.c., f.c.c.,
and b.c.c.\  lattices.
\label{table1}}
\begin{tabular}{ll}
lattice & vectors\\
\tableline
\tableline
s.c. &$(1,0,0), (0,1,0), (0,0,1), (-1,0,0), (0,-1,0), (0,0,-1)$\\
f.c.c. &$(1,1,0), (1,-1,0), (-1,-1,0), (-1,1,0), (1,0,1), (-1,0,1),$\\
&$(1,0,-1), (-1,0,-1), (0,1,1), (0,-1,1), (0,1,-1), (0,-1,-1)$\\
b.c.c.&$(1,1,1), (1,1,-1), (-1,1,1), (-1,1,-1), (-1,-1,1), (-1,-1,-1),
(1,-1,1), (1,-1,-1)$\\
\end{tabular}
\end{table}

\begin{table}
\caption{Values of the universal finite-size correction exponent,
$\Omega$, and the scaling  function exponent, $\sigma$, for the s.c.,
f.c.c., and b.c.c.\  lattices. ($\sigma$  was not found for
the s.c.\ lattice.)
\label{table2}}
\begin{tabular}{lll}
lattice & $\Omega$ & $\sigma$ \\
\tableline
\tableline
s.c. & 0.63 & $\ldots$ \\
f.c.c. & 0.66 & 0.4453 \\
b.c.c.&  0.66 & 0.4433 \\
\end{tabular}
\end{table}

\begin{table}
\caption{Non-universal coefficients for finite-size correction
and the scaling function of the 
s.c., f.c.c., and b.c.c.\  lattices. ($C$ was not found for s.c.\ lattice.)
\label{table3}}
\begin{tabular}{llll}
lattice & $A$ & $B$ & $C$ \\
\tableline
\tableline
s.c. & 0.813 & 0.178 &  $\ldots$\\
f.c.c. &0.767 & 0.186 & 7.95\\
b.c.c.&  0.776 & 0.187 & 5.57 \\
\end{tabular}
\end{table}

\end{document}